\documentclass[twocolumn]{aastex61}
\usepackage{graphicx}
\usepackage{amsmath}
\usepackage{hyperref}
\usepackage{bm}

\begin{document}

\title{The spectral properties of the bright FRB population}
\shorttitle{FRB spectral properties}
\shortauthors{Macquart et al.}

\correspondingauthor{Jean-Pierre Macquart}
\email{J.Macquart@curtin.edu.au}

\author{J.-P. Macquart}
\affiliation{International Centre for Radio Astronomy Research, Curtin University, Bentley, WA 6102, Australia}
\affiliation{ARC Centre of Excellence for All-sky Astrophysics (CAASTRO), Australia}

\author{R.~M. Shannon}
\affiliation{Centre for Astrophysics and Supercomputing, Swinburne University of Technology, PO Box 218, Hawthorn, VIC 3122, Australia}

\author{K.~W. Bannister}
\affiliation{Australia Telescope National Facility, CSIRO Astronomy and Space Science, PO Box 76, Epping, NSW 1710, Australia}

\author{C.~W. James}
\affiliation{International Centre for Radio Astronomy Research, Curtin University, Bentley, WA 6102, Australia}

\author{R.~D. Ekers}
\affiliation{Australia Telescope National Facility, CSIRO Astronomy and Space Science, PO Box 76, Epping, NSW 1710, Australia}
\affiliation{International Centre for Radio Astronomy Research, Curtin University, Bentley, WA 6102, Australia}

\author{J.~D. Bunton}
\affiliation{Australia Telescope National Facility, CSIRO Astronomy and Space Science, PO Box 76, Epping, NSW 1710, Australia}

\begin{abstract}
We examine the spectra of 23 fast radio bursts detected in a fly's-eye survey with the Australian SKA Pathfinder, including those of three bursts not previously reported.  The mean spectral index of $\alpha = -1.6_{-0.2}^{+0.3}$ ($F_\nu \propto \nu^\alpha$) is close to that of the Galactic pulsar population. The sample is dominated by bursts exhibiting a large degree of spectral modulation: 17 exhibit fine-scale spectral modulation with an rms exceeding 50\% of the mean, with decorrelation bandwidths (half-maximum) ranging from $\approx 1$ to $49$ MHz.  Most decorrelation bandwidths are an order of magnitude lower than the $\gtrsim 30\,$MHz expected of Galactic interstellar scintillation at the Galactic latitude of the survey, $|b| = 50 \pm 5 \deg$.  
A test of the amplitude distribution of the spectral fluctuations reveals only 12 bursts consistent at better than a 5\% confidence level with the prediction of 100\%-modulated diffractive scintillation.  Moreover, five of six FRBs with a signal-to-noise ratio exceeding $18$ are consistent with this prediction at less than 1\% confidence.  Nonetheless, there is weak evidence (88-95\% confidence) that the amplitude of the fine-scale spectral modulation is anti-correlated with dispersion measure (DM) that would suggest it originates as a propagation effect.  
This effect appears to be corroborated by the smoothness of the higher-DM Parkes FRBs, and could arise due to quenching of diffractive scintillation (e.g. in the interstellar medium of the host galaxy) by angular broadening in the intergalactic medium.
\end{abstract}


\section{Introduction} \label{sec:intro}

The extreme brightness of the radiation observed from fast radio bursts (FRBs) requires that it is generated by a coherent emission process.  
The millisecond durations of FRB events and their inferred cosmological distances \cite[][]{Lorimeretal2007,Tendulkaretal2017} together imply that their brightness temperatures exceed $10^{35}\,$K, surpassing those of most radio pulsars, whose emission has long been recognised to inescapably involve generation by a coherent process \citep[see][for a recent review]{MelroseYuen2016}.  A yet more acute problem is posed by the luminosity: the $1$--$800$\,Jy flux density emission observed in the bright FRB population, if generated at cosmological distances, is inferred to be over twelve orders of magnitude more luminous than the brightest Galactic pulsars.  Moreover, the $1$--$390\,$Jy\,ms fluences imply energies in excess of $\sim 10^{34}\,(\Omega_b/4 \pi)\,$J for a beam solid angle $\Omega_b$ even if the emission is only confined to the observational bandwidth of $\approx 300\,$MHz  \citep{Bannisteretal2017}.

The origin of such luminous emission is a matter of conjecture \cite[e.g.][]{2008PhRvL.101n1301V,2016MNRAS.457..232C,Metzgeretal2017}.  As such, the spectrum of the emission is both fundamental to characterizing the emission process and to quantifying the total radio energy output.  
Furthermore, fine-scale spectral structure is detected in some FRBs \citep{Ravietal2016,Oslowskietal2018,Farahetal2018} and is even time-variable in FRB121102 \citep{Spitleretal2016}.  


The spectral properties of the FRB population have hitherto been unclear due to the manner in which most of these events have been detected.  The 64-m Parkes radio telescope, responsible for the plurality of FRB detections to date \citep{Petroffetal2016}, uses a 13-beam multibeam receiver whose large separation between adjacent beams renders each burst location sufficiently uncertain that beam chromaticity effects insert substantial, generally indeterminable structure in the shape of the observed spectrum.  By contrast, the phased-array feeds (PAFs) of the Australian SKA Pathfinder ASKAP \cite[ASKAP,][]{McConnelletal2016} enable each burst to be localized sufficiently well to eliminate the  effects of beam chromaticity.


\cite{Shannonetal2018} presented the discovery of $20$~FRBs in a fly's eye survey conducted at a Galactic latitude of $|b| = 50 \pm 5~\deg$ under the auspices of the Commensal Real-time ASKAP Fast Transients \citep[CRAFT;][]{Macquartetal2010} survey on ASKAP.  Here we examine spectral properties of this bright FRB population.  In Section \ref{sec:newfrbs}, we augment the sample with three additional FRB detections. In Section \ref{sec:Results} we analyse the fluence spectra of these ASKAP-CRAFT 23 FRBs.  In Section \ref{sec:Interpretation} we examine several possible interpretations of their remarkable spectral structure, and in Section \ref{sec:Discussion} we discuss the wider implications of our results.

\section{Recent fast radio burst searches}\label{sec:newfrbs}

The analysis presented here is based on the CRAFT observations reported in \cite{Bannisteretal2017}, with further details in \cite{Shannonetal2018}.  Since the conclusion of those searches we have undertaken a few small additional surveys.
Searches were conducted in the same observing band, centered at $1320$~MHz, with a total of $336$~MHz of bandwidth, subdivided into samples of $1$~MHz width and $1.26$\,ms duration.  
Burst properties were measured from the data using the techniques presented in \cite{Bannisteretal2017} and \cite{Shannonetal2018}.

\begin{figure*}[!ht]
\begin{center}
\begin{tabular}{ccc}
  \includegraphics[scale=0.5]{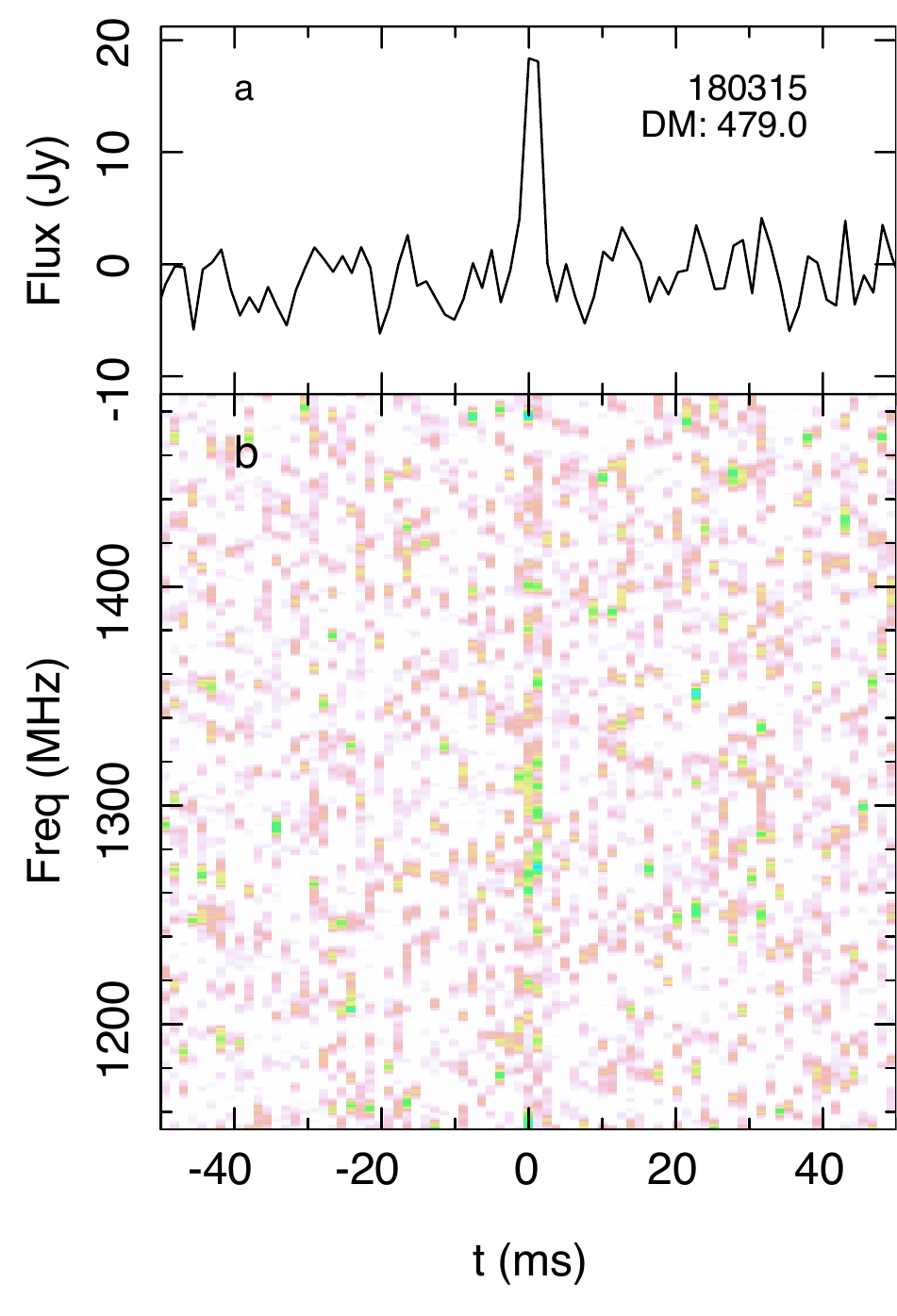} &   \includegraphics[scale=0.5]{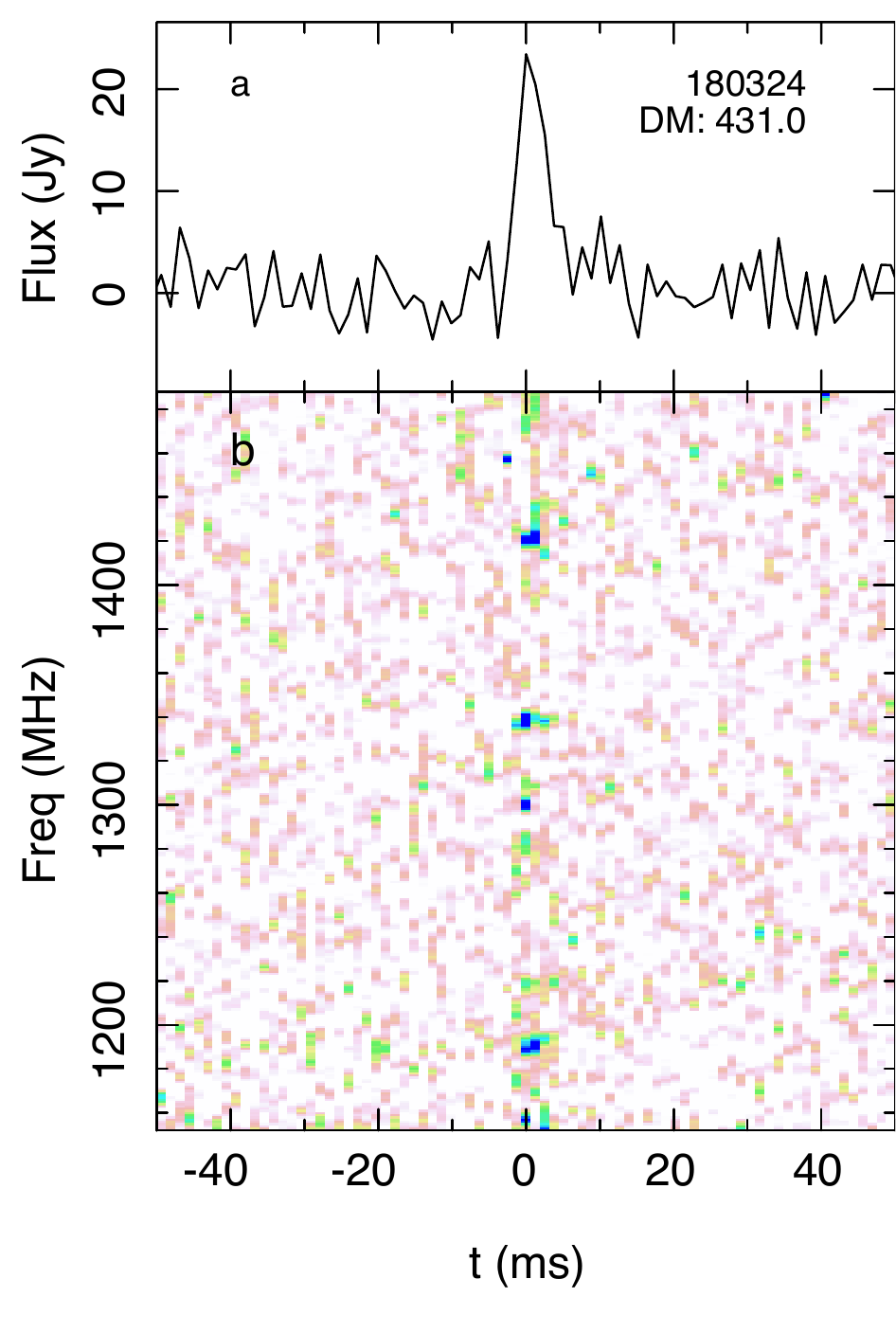}    & \includegraphics[scale=0.5]{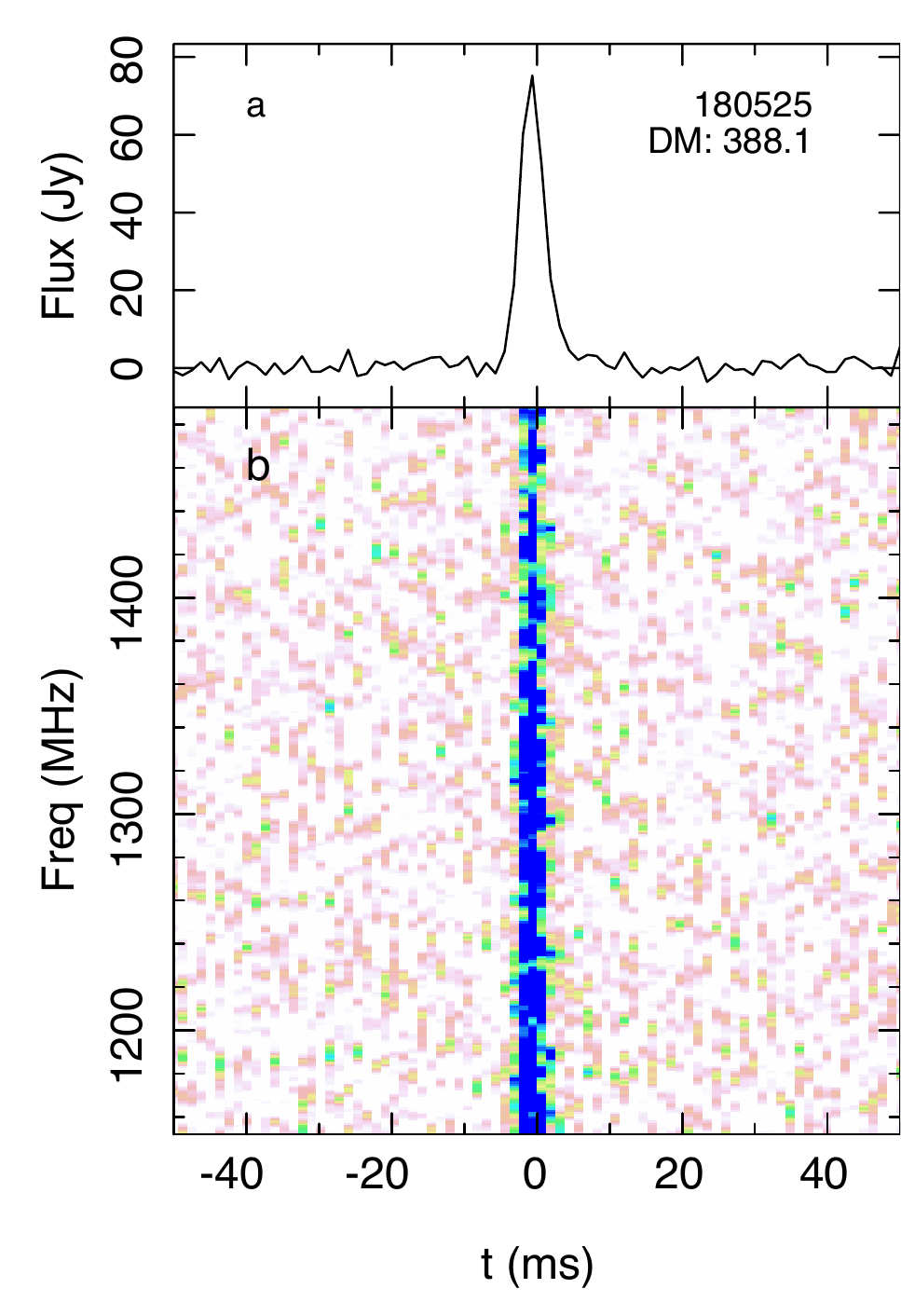} \\
\end{tabular}
\end{center}
\caption{Pulse profiles (panels A) and dynamic spectra (panels B) of newly reported FRBs.  As in Shannon et al. (2018),  the colour scale is set to saturate at $5 \sigma$, where $\sigma$ is the off-pulse rms. 
 \label{fig:newfrbs}  }
\end{figure*}

Here we present the discovery of three additional FRBs in this period.  The properties of the FRBs are listed in Table \ref{tab:newfrbs} and the profiles and spectra are shown in Figure \ref{fig:newfrbs}.

Two FRBs (180315 and 180424) were discovered in a survey conducted at a Galactic latitude of $|b|= 20\deg$.  The Galactic dispersion measure contribution to the bursts  (${\rm DM}_{\rm MW}$, see Table \ref{tab:newfrbs}), the sum of a Galactic disk and halo component, is higher than that of the $b=50^\circ$ FRBs, but still much smaller than the total burst DM.
For the disk component, we have assumed the NE2001 electron density model \cite{NE2001}.  The halo component is estimated to be $15$~pc\,cm$^{-3}$, from the dispersion measure excess in the direction the Large Magellanic cloud FRBs, as elaborated in \cite{Shannonetal2018}.  
The $|b|=20\deg$ searches comprised 71 antenna-days.  Assuming a survey-equivalent effective field of view of $20$~deg$^2$, and $80\%$ observing efficiency \citep{Shannonetal2018} this is a rate of one FRB per 13,600~deg$^2$\,hr exposure, or $72$~sky$^{-1}$\,day$^{-1}$.  
Both show spectral modulation similar to that observed in the sample presented in \cite{Shannonetal2018}. 
The third burst, FRB~180525, was  detected in further high Galactic latitude ($b=50^\circ$) searches.


As our searches frequently re-observed the same fields, we can place strong constraints on repetition. Table \ref{tab:newfrbs} also lists the amount of time observing these fields.  More than $22$~d has been observed in the direction of FRB180525.

\begin{deluxetable*}{lllrllrrrrrrr}
\tabletypesize{\scriptsize}
\tablehead{}
\tablecaption{Properties of the newly discovered FRBs \label{tab:newfrbs}} 
\startdata
  \multicolumn{1}{c}{FRB} & \multicolumn{1}{c}{Time}  & \multicolumn{1}{c}{DM} &\multicolumn{1}{c}{$E_\nu$} & \multicolumn{1}{c}{R.A. (J2000)}  & \multicolumn{1}{c}{Dec. (J2000)}  & \multicolumn{1}{c}{$g_l$} &   \multicolumn{1}{c}{$g_b$}  & \multicolumn{1}{c}{$w$} & \multicolumn{1}{c}{S/N$^{(3)}$}   & \multicolumn{1}{c}{DM$_{\rm MW} $}    & \multicolumn{1}{c}{DM$_{\rm EG} $}  & \multicolumn{1}{c}{$T_{\rm obs}$}    \\
  &   \multicolumn{1}{c}{(TAI$^{(1)}$)} & \multicolumn{1}{c}{(pc\,cm$^{-3}$)} &  \multicolumn{1}{c}{(Jy\,ms) }& \multicolumn{1}{c}{(hh:mm)$^{(2)}$} & \multicolumn{1}{c}{(dd:mm)$^{(2)}$} & \multicolumn{1}{c}{(deg.)} & \multicolumn{1}{c}{(deg.)}  & \multicolumn{1}{c}{(ms)} & \multicolumn{1}{c}{}   & \multicolumn{1}{c}{(pc\,cm$^{-3}$)} & \multicolumn{1}{c}{(pc\,cm$^{-3}$)}  & \multicolumn{1}{c}{(d)}\\
\hline
180315 &05:06:07.9851(2)&  479.0(4) &   56(4) &  19:35(3) & $-$26:50(10) & 13.2 & $-$20.9 & 2.4(3) & 10.4 & 116 & 363 & 2.5\\
180324 & 09:32:23.7066(3) & 431.0(4) &   71(3) & 06:16(3) & $-$34:47(10) & 245.2 & $-$20.5  &  4.3(5) & 9.8 & 79 & 352 &  2.0 \\
180525 &  15:19:43.51508(3) & 388.1(3) &   300(6)  & 	14:40(2) &$-$02:12(6)  & 349.0 & 50.7 & 3.8(1) & 27.4 & 46 &  303 & 21.6  \\
\enddata
\tablecomments{  (1)  Arrival times refer to TAI (not UTC) and are referenced to a frequency of $1297$~MHz.  (2)  Uncertainties are the marginalized 90\% containment regions.  (3) S/N as reported in the search.   (4)  The Milky-Way DM (DM$_{\rm MW}$)  contribution is the sum of a disk \cite[][]{NE2001} and halo DM components; the extra-Galactic DM component (DM$_{\rm EG}$) is the difference between the two.  } \label{tab:FRBtable}
\end{deluxetable*}

\begin{figure}[!ht]
\begin{center}
\begin{tabular}{c}
  \includegraphics[scale=0.6]{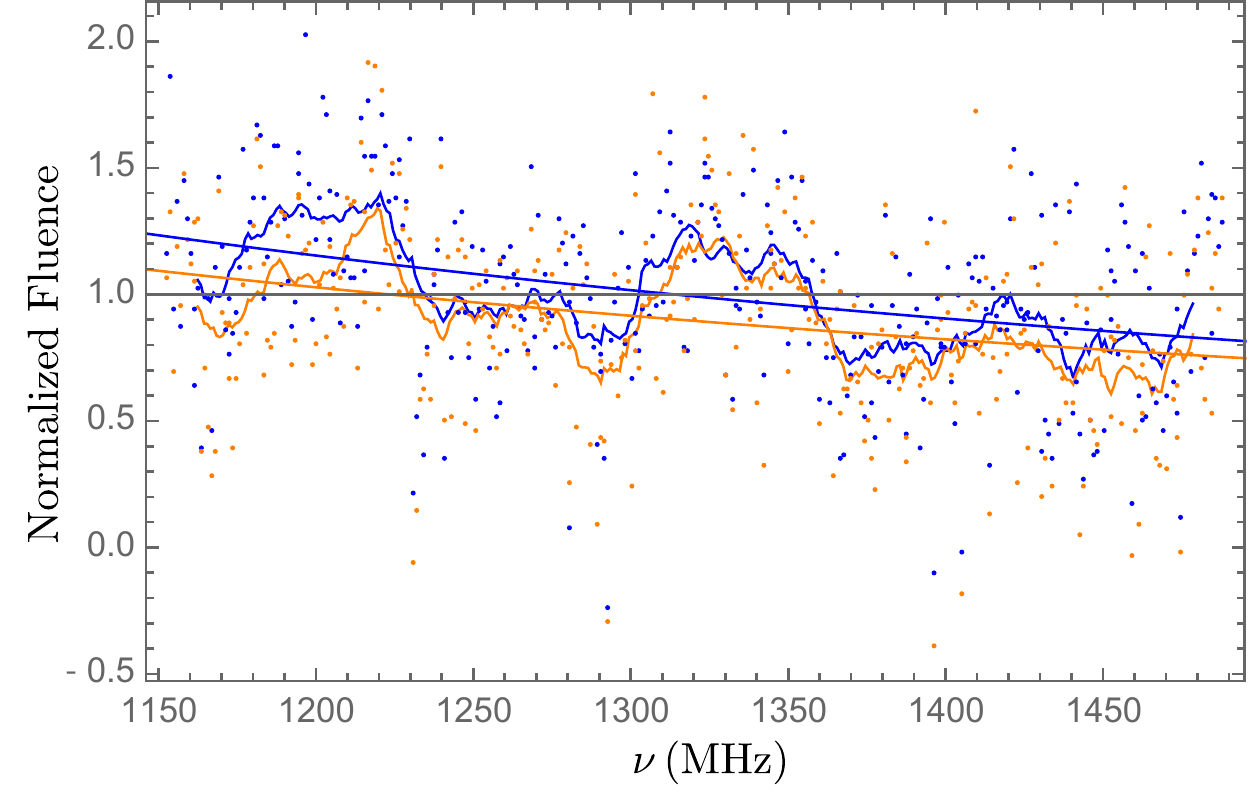}
\end{tabular}
\end{center}
\caption{The equal-weight average flux density spectrum from the set of 23 ASKAP-CRAFT FRBs. Blue and orange points indicate, respectively, the mean and median fluence of each spectral channel, the light blue and orange curves indicate the 20-MHz moving average of the mean and median fluence, and the heavy lines indicate their corresponding best-fit power law curves.  
 \label{fig:mean_spectrum}  }
\end{figure}

\begin{figure*}[!ht]
\begin{center}
\begin{tabular}{cc}
\includegraphics[scale=0.3]{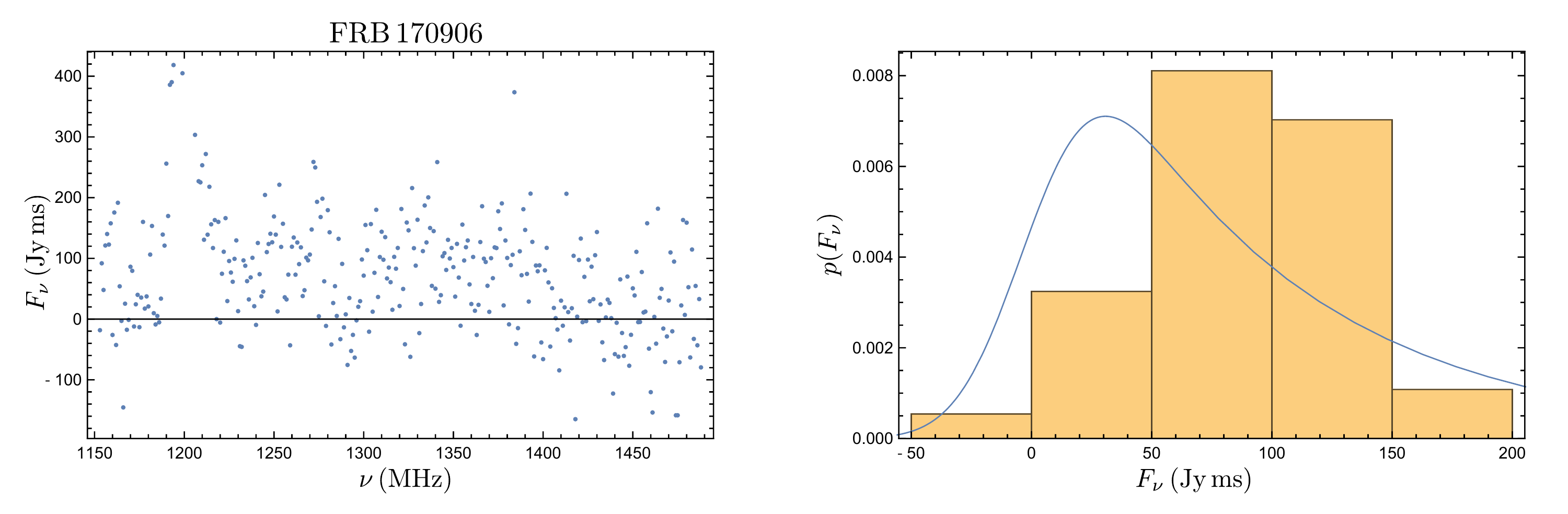} & \includegraphics[scale=0.3]{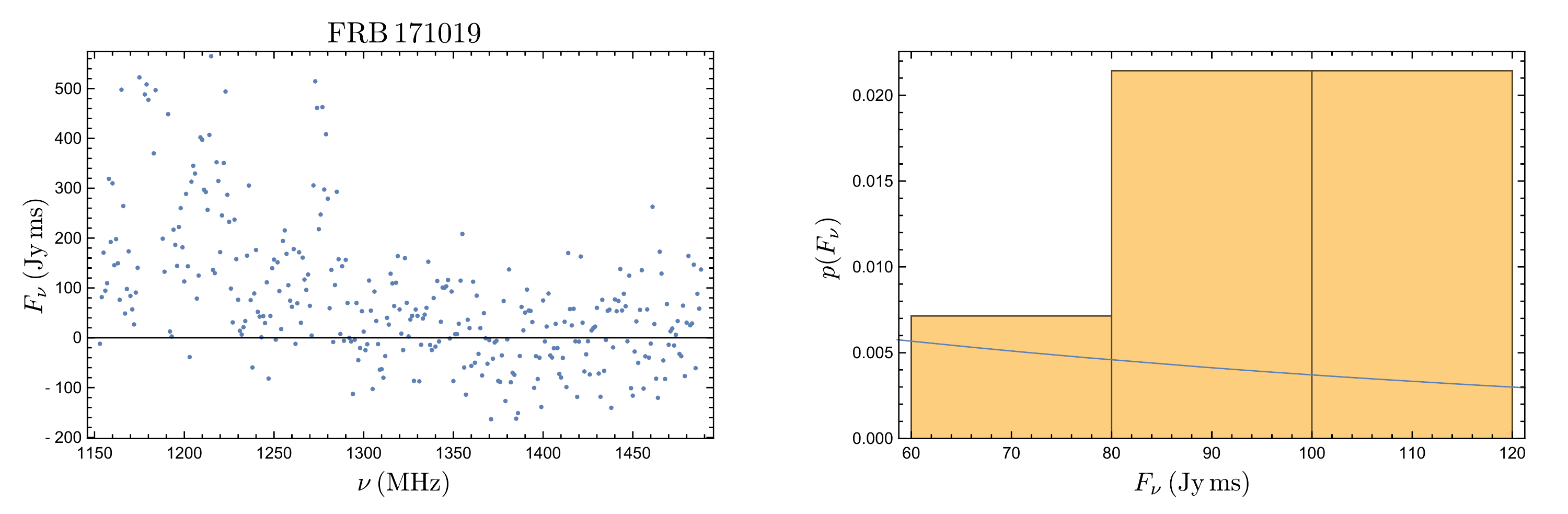} \\
\includegraphics[scale=0.3]{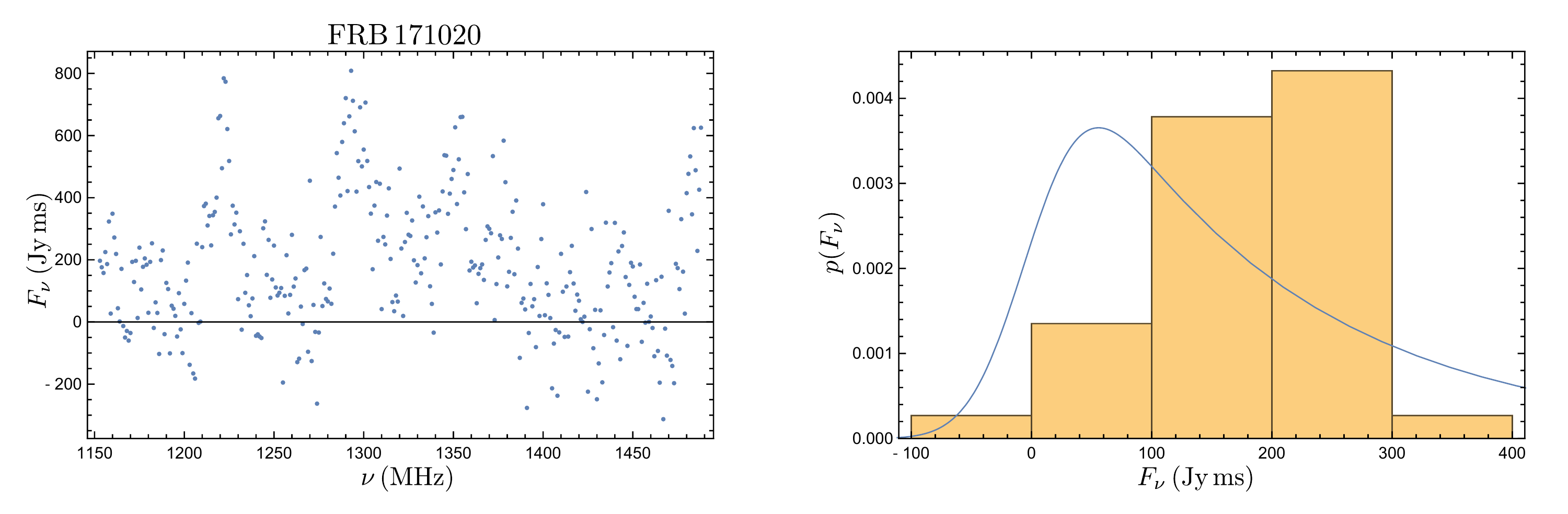} & \includegraphics[scale=0.3]{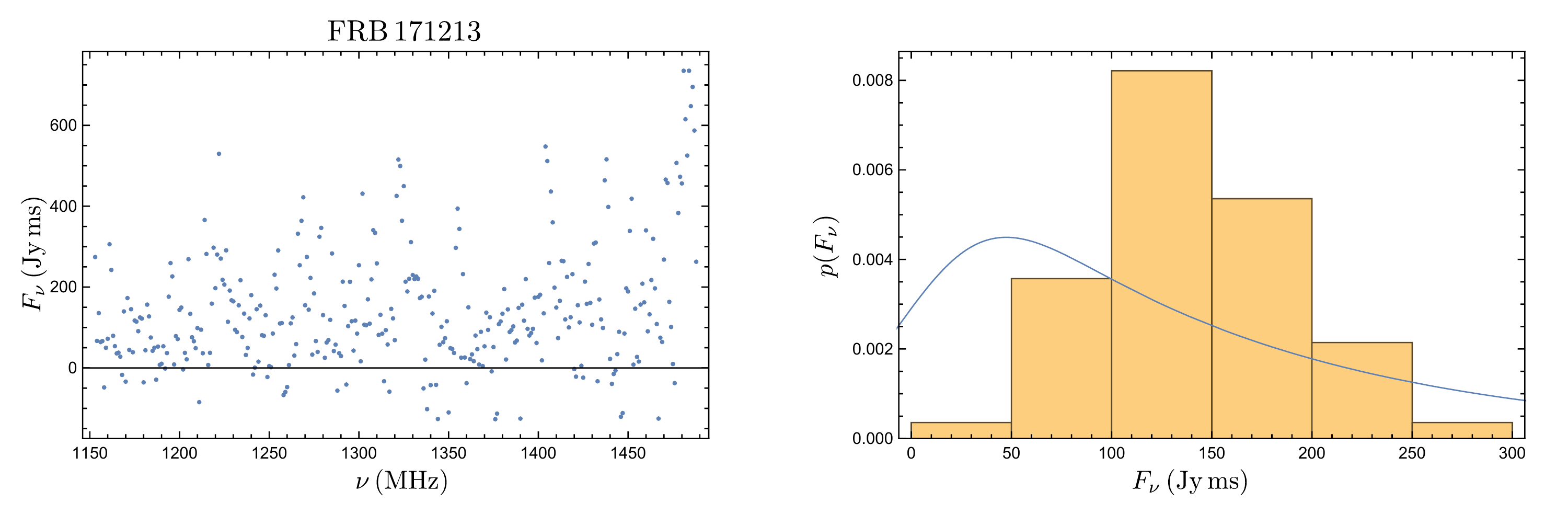} \\
\includegraphics[scale=0.3]{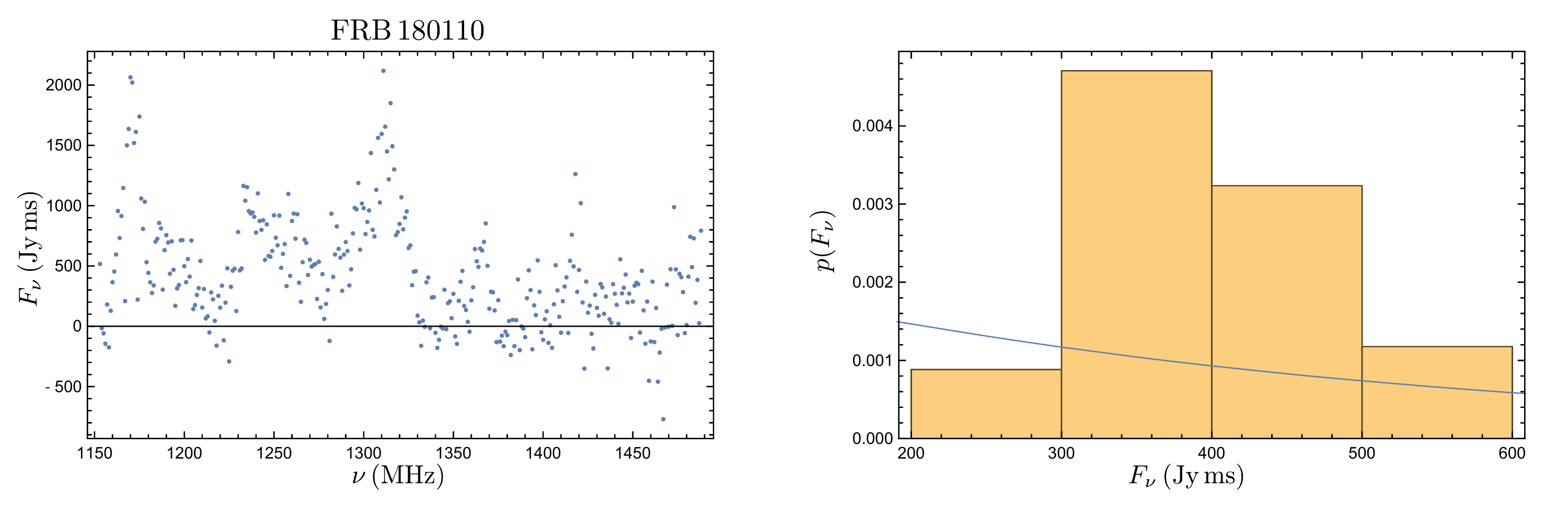} & \includegraphics[scale=0.3]{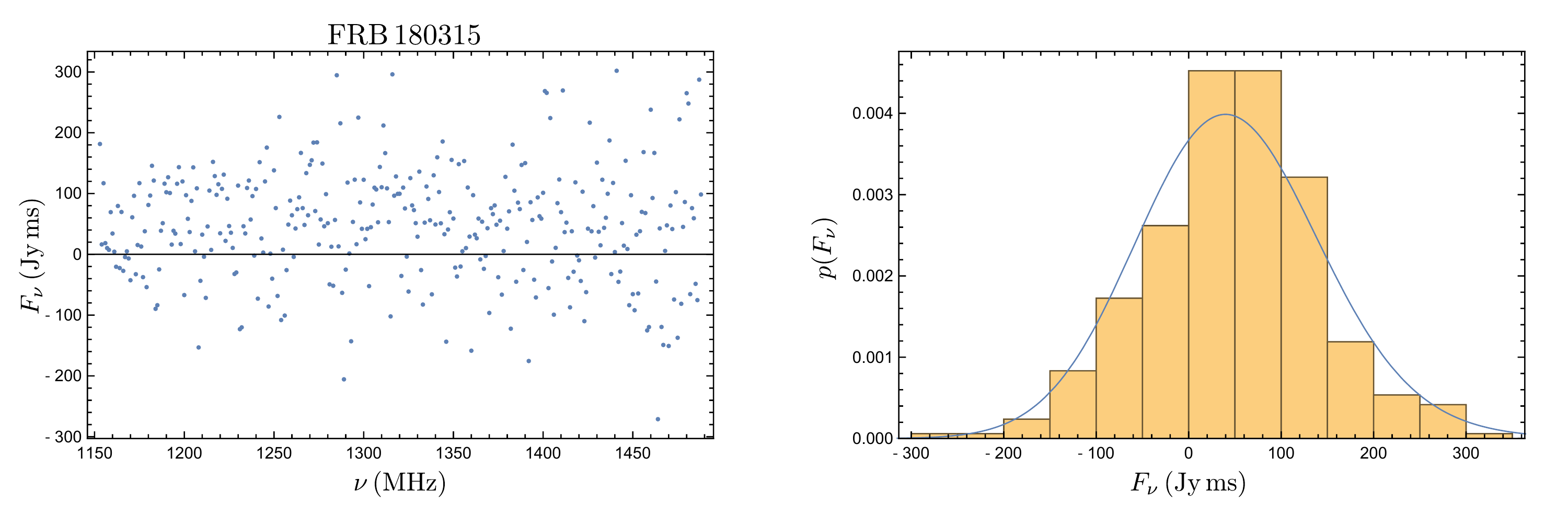} \\
\includegraphics[scale=0.3]{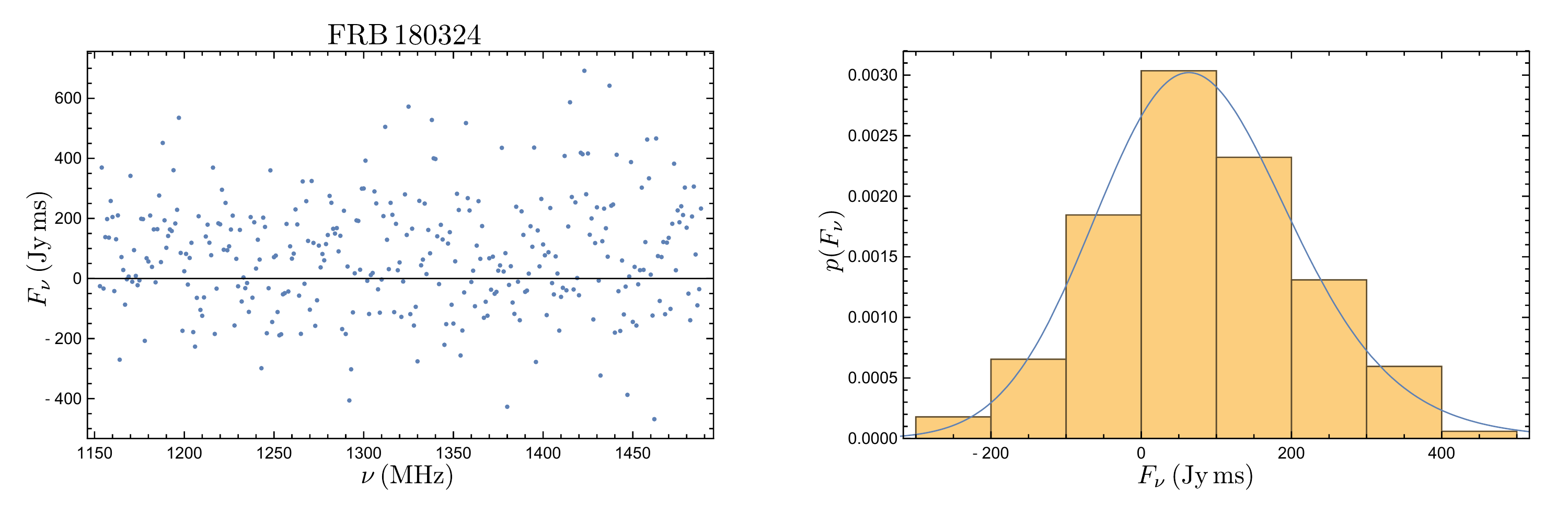} & \includegraphics[scale=0.3]{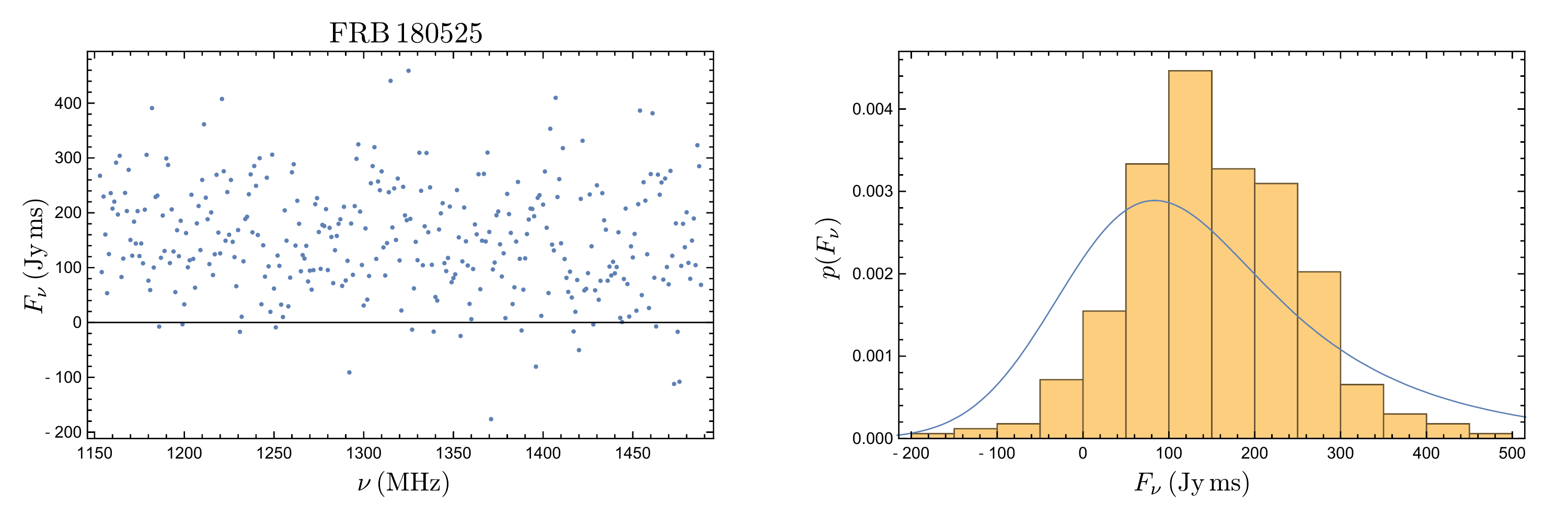} \\
\end{tabular}
\end{center}
\caption{Spectra (left) and the corresponding fluence histograms of all bursts with $S/N > 18$ (frames 1-5 and 8) and the newly-detected bursts (frames 6-8), after binning to a spectral resolution $\nu_{\rm dc}$ (right).  
The blue curve shows the predicted probability distribution function for fully-modulated diffractive scintillation, as shown in eq.(\ref{DiffDistn}).
} \label{fig:FRBDistributionFitTest}
\end{figure*}

\section{Observations and Spectral Properties} \label{sec:Results}
The spectra reported here were corrected for the variation in $T_{\rm sys}$ across the observing band in the same manner reported in \cite{Shannonetal2018}.   The close spacing of beams within the ASKAP PAFs enables each burst to be localized to within a small fraction of the beamwidth, typically within a region $10\arcmin \times 10\arcmin$.  The spectrum and fluence of each burst were corrected for the small off-axis beam chromaticity using the fact that, for the small offsets from beam center relevant to these FRBs, the beam is well approximated as a Gaussian with a full-width at half power of $1.1 c/(\nu D)\,$rad, where $D=12\,$m is the dish diameter \citep{McConnelletal2016}. Both FRB\,171216 and FRB\,180525 were detected with comparable S/N in two beams, with offsets of $0.491^\circ$ and $0.43^\circ$, and $0.5^\circ$ and $0.513^\circ$, respectively from the beam center.  We applied the beam correction to the spectrum of each beam separately before computing the weighted average spectrum of the data from the two beams.  


\subsection{Broadband spectrum}

Figure \ref{fig:mean_spectrum} shows the mean and median spectrum of the 23 FRBs in our sample, with each burst given equal weight in the average by normalizing by its mean fluence.  A strong trend of decreasing fluence with frequency is visible over the 336\,MHz observing band, which  we characterize by fitting a power law of the form $F_\nu \propto \nu^\alpha$.  The spectral index of the mean spectrum is $\alpha = -1.6_{-0.2}^{+0.3}$, while for the median spectrum it is $\alpha = -1.4_{-0.3}^{+0.3}$.  
These values are within 1 standard deviation of the results reported in \cite{Shannonetal2018} on the basis of the first 20 FRBs detected in the CRAFT survey. 

Table \ref{tab:SpectrumTable} lists the spectral indices obtained from the fits to individual spectra.  The median spectral index is $\alpha=-1.1$. The large variation in $\alpha$ is in part a reflection of the patchiness of the emission.   For instance, for FRBs 170906 and 171019 the spectrum is entirely dominated by emission in the lower half of the band, whereas for FRB\,180128.2 the spectrum is dominated by the upper half of the band.  

Although there is large variation in individual burst properties, we verified that the mean spectrum measured from this set of 23 bursts converges to a physically meaningful quantity.  
We addressed the issue of convergence to a population spectral index by measuring the mean spectral index measured from all subsets consisting of $N=3, 4, \ldots, 22$ bursts.  
The spectral index computed from the mean over all $N$-length subsets is found to converge as $N$ increases to the measured value, with a decrease in the standard deviation of the spectral index distribution that scales approximately $\propto N^{-1/2}$.

\subsection{Characterization of spectral variations} \label{sec:CharacterVars}
A significant subset of the CRAFT bursts exhibit large amplitude narrowband spectral variability.  We attempt to estimate the amplitude of the variability using two approaches: (i) $m^2$: we compute the mean-normalized spectral autocovariance directly from the spectrum, $f(\Delta \nu) = \langle [F_\nu (\nu'+\Delta \nu)-\bar{F_\nu}] [F_\nu (\nu') -\bar{F_\nu} ] \rangle/\bar{F_\nu}^2$, in the limit $\Delta \nu \rightarrow 0$; (ii) $m_{\alpha,{\rm corr}}^2$: we compute the same limit, but with $\bar{F_\nu}$ replaced by the best-fitting function of the form $K \nu^{-1.5}$ whose value $K$ is chosen to preserve the band-average fluence of the burst. 
The decorrelation bandwidth, $\nu_{\rm dc}$ (half width at half power) is also computed from $f(\Delta \nu)$ in each case.
Both methods were further checked for consistency against a robust estimate of the variance by subtracting in quadrature the variance due to the thermal noise, measured from the off-pulse spectrum, from the total spectral variance. The 32/27 oversampling of the 1\,MHz coarse spectral channels on ASKAP results in at most a correction of 1\% to the values of $m^2$ we report here.


A possible cause of this narrowband structure is diffractive scintillation, which makes a clear prediction for the statistics of the intensity fluctuations.  
The amplitude distribution of a point source subject to diffractive scintillation is an exponential \citep{Mercier1962,Salpeter67} which, when convolved with normally-distributed\footnote{In the present case, noise in the dynamic spectrum is expected to follow a $\chi^2$ distribution with $> 3000$ degrees of freedom which is excellently approximated by a normal distribution.} thermal noise, ${\cal N}(I)$ with variance $\sigma_t$, predicts spectral fluctuations of the form:    
\begin{eqnarray}
p_I (I) &=& \int_0^\infty \frac{1}{I_0} \exp \left( - \frac{I'}{I_0} \right) {\cal N}(I - I') dI'  \nonumber \\
&=& \frac{1}{2 I_0} \exp \left( \frac{\sigma_t^2}{2 I_0^2} - \frac{I}{ I_0} \right) {\rm erfc} \left[  \frac{1}{\sqrt{2}} \left( \frac{\sigma_t}{ I_0} - 
\frac{I}{\sigma_t} \right) \right], \nonumber \\  \label{DiffDistn}
\end{eqnarray}
where erfc is the complementary error function, and $I_0$ is the mean flux density. Since the scintillation signal is highly correlated over a spectral range $\approx \nu_{\rm dc}$, we  test this distribution against the statistics of each spectrum binned to a resolution of $\lceil \nu_{\rm dc} \rceil$. The rms noise, $\sigma_t$, is measured directly from off-pulse data adjacent to the burst arrival time, and scaled according to the number of 1-MHz spectral bins averaged together.  


Table \ref{tab:SpectrumTable} shows the confidence, using the Cram\'er-von Mises test, that the spectral data matches the model in eq.(\ref{DiffDistn}). The test is performed by: (i) comparing the distribution of channelised fluences directly against the distribution, or (ii) using the channelised fluences normalised by the power law whose spectral index matches that of the median burst spectrum (i.e. $\alpha = -1.5$) and whose band-averaged fluence matches that of the burst.  

Twelve bursts have intensity distributions consistent with the null hypothesis of being drawn from eq.(\ref{DiffDistn}) at the 5\% confidence level or greater.  
Figure \ref{fig:FRBDistributionFitTest} shows the observed and predicted distributions for a number of the eleven cases in which the model is an exceptionally poor description of the spectral data.
There is a clear trend for these to be strong events for which the test has the greatest discriminating power:
five of six FRBs with a signal-to-noise (S/N) ratio greater than $18$ are consistent at less than 1\% confidence. The exception, FRB 171019, has $\nu_{\rm dc}=49$\,MHz, and hence the test is weakened by having only seven effective samples across the band.

%
%
%
%
%


Finally, we remark that no obvious absorption features are seen in any of the spectra.
\begin{deluxetable*}{lrrr|cc|cc}
\tabletypesize{\scriptsize}
\tablehead{}
\tablecaption{Spectral properties of individual FRBs \label{tab:SpectrumTable}}
\startdata
 & & & & \multicolumn{2}{|c|}{Unnormalized spectrum} & \multicolumn{2}{|c}{Normalized spectrum} \\
  \multicolumn{1}{c}{FRB} & \multicolumn{1}{c}{S/N (1)} & \multicolumn{1}{c}{$\alpha$}  & \multicolumn{1}{c}{$\nu_{\rm dc}$} & \multicolumn{1}{|c}{$m^2$ (2)} & \multicolumn{1}{c|}{\% confidence fits}  & \multicolumn{1}{|c}{$m_{\alpha,{\rm corr}}^2$ (2)} & \multicolumn{1}{c}{\% confidence fits}  \\
  &  &  &  \multicolumn{1}{c}{MHz}  & \multicolumn{1}{|c}{} &  \multicolumn{1}{c|}{exp. distn.} & \multicolumn{1}{|c}{} & \multicolumn{1}{c}{exp. distn.} \\
\hline
170107 & 13.5 & $-3.5$ & (3.6) & 0.15 & 3 & 0.10  & 5 \\
170416 & 10.4 & $-7.5$ & 15 & 1.0 & 5 & 0.80  & 8 \\ 
170428 & 10.2 & $-2.1$ & --- & 0 (-0.24) & 11 & 0 (-0.21)  & 17 \\ 
170707 & 11.4 & $-1.8$ & 1.0 & 0.49  & 78 & 0.58  & 92 \\ 
170712 & 15.4 & $-4.2$ & 6.9 & 0.39 & 0.3 & 0.33  & 0.4 \\ 
170906 & 25.8 & $-6.3$ & 9.4 & 1.7 & 0.5 & 1.2  & 0.4 \\ 
171003 & 14.0 & $+7.0$ & 48 & 0.48 & 37 & 0.75  & 9 \\ 
171004 & 12.4 & $+3.0$ & 23 & 0.64 & 30 & 0.81 & 32 \\ 
171019 & 23.4 & $-13$ & 49 & 2.3 & 3 & 1.5  & 3 \\ 
171020 & 28.4 & $-0.68$ & 8.4 & 0.86 & 0.08 & 0.87 & 0.09 \\ 
171116 & 11.5 & $+2.0$ & 1.2 & 0.08 & 38 & 0.13 & 28 \\ 
171213 & 28.5 & $+4.4$ & 5.4 & 0.66 & $4 \times 10^{-4}$ & 0.93 & $1 \times 10^{-3}$ \\ 
171216 & 8.8 & $+2.8$ & 2.1 & 1.06 & 95 & 1.05 & 81 \\
180110 & 37.4 & $-4.6$ & 10 & 0.76 & $1 \times 10^{-3}$ & 0.64 & $1 \times 10^{-3}$ \\ 
180119 & 18.3 & $-0.89$ & 1.8 & 0.97 & 86 & 0.89  & 77 \\ 
180128.2 & 12.7 & $+7.5$ & 25 & 0.76 & 5 & 0.82 & 5  \\ 
180128.0 & 15.0 & $-2.2$ & 12 & 0.68 & 0.6 & 0.69 & 1 \\ 
180130 & 11.5 & $+0.76$ & 0.4 & 0.30 & 55 & 0.41 & 50 \\
180131 & 14.4 & $-2.2$ & 1.3  & 0.33 & 1 & 0.32  & 1  \\
180212 & 17.7 & $-3.6$ & 2.8 & 0.60 & 0.3 & 0.51 & 0.2\\
180315 & 9.3 & $-0.52$ & 5.9 & 0.08 & 11 & 0.05 & 8 \\
180324 & 9.0 & $+1.1$ & 1.3 & 0.72 & 93 & 0.84 & 96 \\
180525 & 31.2 &  $-1.1$ & 0.36 & 0.06 & $2 \times 10^{-6}$ & 0.06  & $3 \times 10^{-6}$ \\
\enddata
\tablecomments{(1) This post-detection S/N differs slightly from the detection S/N reported in \cite{Shannonetal2018} and, for FRBs 171216 and 180525, is the result of co-addition of data from two adjacent beams. (2) The measured value of the autocovariance at $\Delta \nu=1\,$MHz is negative for FRB~170428, so we take $m^2 = m_{\alpha,{\rm corr}}^2=0$ for this burst.  However, we note that this could instead indicate spectral modulation with $\nu_{\rm dc}\ll 1\,$MHz, in which case the data could be more consistent with a diffractive scintillation interpretation at better than the 11-17\% level indicated for this burst.}
\end{deluxetable*} 

\section{Interpretation} \label{sec:Interpretation}

The most significant feature of the bright burst spectra observed by ASKAP is the high degree of spectral modulation. 
It is an open question whether these are characteristic of the burst emission process or of a propagation effect. The former is expected on the grounds that many coherent emission processes (e.g.\,pulsars, Jovian decametric radiation, solar radio bursts) exhibit fine spectral structure, many of which may persist in time \citep[][and references therein]{HankinsEilek2007,Ellis1969,Melrose2017}.  However, the millisecond duration of FRBs indicates they are sufficiently compact that their spectra should also be subject to lensing and diffractive scintillation effects caused by inhomogeneous plasma in our Galaxy, the host galaxy, or possibly the IGM. 

It is far from conclusive to which effect the structure observed in the present sample should be ascribed.  The spectral structure might be associated with caustics due to plasma lensing, as suggested by \cite{Cordesetal2017}.  However, the complicated spectral structure observed in many bursts does not generically resemble the predictions of \cite{Cordesetal2017} wherein only one or two sharply-peaked caustics are expected in the frequency domain, rather than the large number of bright islands of power evident in most spectra (the clearest examples being FRBs 170416, 171019, 180110 and 180324).  Moreover, the spectra do not qualitatively resemble those of other astrophysical events recognized to be associated with caustics, notably extreme scattering events \citep[e.g.][]{Bannisteretal2016}.


Diffractive scintillation provides a more viable interpretation of the narrowband structure.  However, for the high Galactic latitude ($|b| \approx 50^\circ$) FRBs reported here, the decorrelation bandwidth expected due to Galactic interstellar scintillation is $\sim 20$--$200\,$MHz \citep{NE2001,Bhatetal2018}; all but three bursts in the sample exhibit decorrelation bandwidths less than 20\,MHz (although the smooth spectrum of FRB\,180525 might be caused by an extremely broadband scintle).   Thus, if the structure is due to scintillation, it is more readily attributed to a medium external to the Galaxy: either in the interstellar medium of the burst host galaxy, or in some overdense region of intergalactic plasma along the line of sight.  

If the diffractive scintillation model fits the data at all it is clearly largely inconsistent with the simple prediction that the intensity fluctuations follow an exponential distribution.
Table \ref{tab:SpectrumTable} shows that over half of the bursts are inconsistent with fully-modulated diffractive scintillation, and none for which there is good discriminating power (high S/N, low $\nu_{\rm dc}$), as embodied in eq.\,(\ref{DiffDistn}).  That high S/N events over all DMs show strong evidence against the diffractive scintillation model suggests there is also no correlation with FRB burst energy. 

There is nonetheless weak evidence for an anti-correlation between the DM and the amplitude of the spectral modulation, which would suggest the spectral structure does arise as a propagation effect. We examined the likelihood of the hypothesis that the ${\rm DM}_{\rm EG}$ and $m_{\alpha,{\rm corr}}$ data are independent in the ASKAP-CRAFT sample.  The Spearman rank test and Kendall tau tests return likelihoods of 6\% and 8\%, with correlation coefficients of $-0.4$ and $-0.3$ respectively.  The Pearson correlation test returns a likelihood of 5\%.

There are three obvious ways in which the spectral structure may still arise as a result of diffractive scintillation: the number of scintles sampled across the observing band is finite, the structure is spectrally unresolved, or the diffractive scintillation is partially quenched by some physical process.   The first explanation is unlikely, since the statistical test used to calculate the confidence values in Table \ref{tab:SpectrumTable} takes into account the finite number of samples.  

Of the second and third possibilities, it is remarkable that the Parkes FRBs, located at a higher average DM, exhibit generally smoother spectra \citep{Thorntonetal2013,Championetal2016}.  Moreover, two of the lowest DM  Parkes bursts \cite[FRBs 150807 and 180309, ][]{Ravietal2016,Oslowskietal2018} and the lowest DM burst detected by UTMOST \cite[FRB 170827][]{Farahetal2018} do show strong spectral modulation consistent with the ASKAP-CRAFT bursts.   

There are two obvious ways in which an anti-correlation between DM and the amplitude of the spectral modulation could arise if the structure were propagation-induced:  

1. DM scales with the scattering measure, causing a commensurate decrease in $\nu_{\rm dc}$.  For sufficiently strong scattering $\nu_{\rm dc}$ would fall below our 1\,MHz spectral resolution. However, the data disfavor this explanation because in most cases the spectral structure is either resolved or marginally resolved.  Moreover, suppression of the modulations to $m^2 < 0.1$ would require $\nu_{\rm dc} \lesssim 100\,$kHz. This would require $\nu_{\rm dc}$ in these bursts to be considerably lower than the lowest observed at 1.4\,GHz with higher-resolution instruments: the narrowest structure at 1.4\,GHz has $\nu_{\rm dc} = 100 \pm 50\,$kHz \citep[in FRB150807;][]{Ravietal2016}.

2. Angular broadening partially suppresses the amplitude of the spectral variations associated with diffractive scintillation.   The \cite{Shannonetal2018} DM-fluence relation shows that the FRB DMs are related to distance, and the only component of the DM that relates to burst distance is the IGM contribution.  A correlation between angular broadening and DM is expected if, as higher DM implies greater distance, there is a greater likelihood of grazing the halo of an intervening galaxy along the light of sight harboring plasma sufficiently dense to produce an appreciable amount of angular broadening.

The optics of angular broadening heavily weights the contribution from scattering material closer to the observer, so the amount of angular broadening need not be directly coupled with temporal smearing, which favors the contribution of material closer to mid-way along the ray-path.  Thus the absence of temporal smearing \citep[e.g. in the ${\rm DM}=2596\,$pc\,cm$^{-3}$ FRB160102;][]{Bhandarietal2018} does not imply an absence of scatter broadening and, conversely, the presence of temporal smearing (e.g. in FRB180110) does not necessarily imply angular broadening.




\section{Discussion} \label{sec:Discussion}
The mean spectral index measured for the bright FRB population, $\alpha = -1.6_{-0.2}^{+0.3}$, is consistent with the range $\alpha=-1.4$ to $-1.6$ typically derived for the slow and millisecond pulsar populations \citep{Batesetal2013,Jankowskietal2018}, but apparently at variance with the spectral index of giant pulses and magnetars.  An index of $\alpha = -2.6$ is derived from the giant pulses from the well-studied Crab pulsar between 0.7 and 3.1\,GHz \citep{Meyersetal2017}, while the range of radio spectral indices typical of magnetar emission is flat, with $\alpha > -0.5$ \citep{Camiloetal2007,Camiloetal2008,Levinetal2010,Eatoughetal2013}.  Although it is premature to draw conclusions on the mechanism responsible for the ultra-luminous emission from FRBs, the similarity in spectral indices hints at an association with spin-down-powered pulsar emission.

The steep spectral index of FRB emission implies that the location of the low-frequency turnover is crucial in understanding the total radio energy output.  The 200\,MHz MWA non-detections of several bright FRBs in the sample analyzed here sets limits on the total radiative output below 1.5\,GHz to $<18$ times that observed in the ASKAP band \citep{Sokolowksietal2018}. However, the reported CHIME detections of some FRBs above 400\,MHz \citep{Boyleetal2018} appear to bracket the range of any spectral turnover.

The steepness of bright FRB emission underlines the importance of the k-correction in the determination of burst luminosities and energies 
\citep[see the discussion in][]{MacquartEkers2018b}. For a burst at luminosity distance $D_L(z)$ the fluence and intrinsic energy density, $E_\nu$ are related by $F_\nu = (1+z)^{2+\alpha} E_\nu/(4 \pi D_L^2(z))$.  Thus the fluence of a $z=1$ burst is a factor of $2.8$ lower relative to one of same energy density but a flat spectrum.

A pervasive feature of the ASKAP-CRAFT FRB spectra is their patchiness.  There is a large degree of spectral variation between individual bursts, resembling the spectrally erratic emission from the repeating FRB 121102 \citep{Spitleretal2016,scholz16}.  

The origin of this spectral structure in the bright FRB population remains inconclusive, and compelling reasons remain to suggest that the structure may yet be an intrinsic property of the bursts.  Nonetheless, if the inverse relation between the amount of spectral structure and the DM weakly favored by the ASKAP-CRAFT data should be confirmed, we suggest it may be caused by the angular broadening in the halos of intervening galaxies embedded in the IGM.

The generally smoother spectral structure observed in Parkes FRBs lends credence to the hypothesis that the fine spectral structure is indeed a propagation effect. Incorporation of spectral information from the full Parkes FRB sample would provide a far stronger statistical test for the presence of a DM-spectral modulation relation.


\acknowledgments

The Australian SKA Pathfinder is part of the Australia Telescope National Facility which is managed by CSIRO. Operation of ASKAP is funded by the Australian Government with support from the National Collaborative Research Infrastructure Strategy. ASKAP uses the resources of the Pawsey Supercomputing Centre. Establishment of ASKAP, the Murchison Radio-astronomy Observatory and the Pawsey Supercomputing Centre are initiatives of the Australian Government, with support from the Government of Western Australia and the Science and Industry Endowment Fund. We acknowledge the Wajarri Yamatji people as the traditional owners of the Observatory site.  JPM, RMS and KB acknowledge funding through Australian Research Council grant DP180100857.    R.M.S. also acknowledges support through Australian Research Council (ARC) grants FL150100148 and CE17010000.



\bibliographystyle{aasjournal}
\bibliography{references-CRAFT_Spectra}


\end{document}